\documentstyle[epsf,twoside,ijmpc1]{article}

\def\qed{\hbox{${\vcenter{\vbox{			
   \hrule height 0.4pt\hbox{\vrule width 0.4pt height 6pt
   \kern5pt\vrule width 0.4pt}\hrule height 0.4pt}}}$}}

\def\bsc{{\sc a\kern-6.4pt\sc a\kern-6.4pt\sc a}}  	
\def\bflatex{\bf L\kern-.30em\raise.3ex\hbox{\bsc}\kern-.14em 
T\kern-.1667em\lower.7ex\hbox{E}\kern-.125em X} 

\begin{document}

\runninghead{Effect of shear on droplets in a binary mixture} 
{Alexander J. Wagner \& J. M. Yeomans}

\normalsize\textlineskip
\thispagestyle{empty}
\setcounter{page}{1}

\copyrightheading{}			

\vspace*{0.88truein}

\fpage{1}
\centerline{\bf EFFECT OF SHEAR ON DROPLETS }
\vspace*{0.035truein}
\centerline{\bf IN A BINARY MIXTURE}
\vspace*{0.37truein}
\centerline{\footnotesize ALEXANDER J. WAGNER}
\vspace*{10pt}
\centerline{\normalsize and}
\vspace*{10pt}
\centerline{\footnotesize J. M. YEOMANS}
\vspace*{0.015truein}
\centerline{\footnotesize\it Theoretical Physics, University
of Oxford, 1 Keble Rd.}
\baselineskip=10pt
\centerline{\footnotesize\it Oxford OX1 3NP, United Kingdom}
\vspace*{0.225truein}
\publisher{(\today)}{(revised date)}

\vspace*{0.21truein} \abstracts { In this article we use a
lattice-Boltzmann simulation to examine the effects of shear flow on a
equilibrium droplet in a phase separated binary mixture. We find that
large drops break up as the shear is increased but small drops
dissolve. We also show how the tip-streaming, observed
for deformed drops, leads to a state of dynamic equilibrium.}{}{}



\vspace*{1pt}\textlineskip	
\section{Introduction}		
\vspace*{-0.5pt}
\noindent
The study of phase transitions under shear has been of great interest
in the recent past\cite{olson,jasnov,goldburg,lauger}. To understand
the basic phenomena underlying this complex process we focus our
attention on the behavior of a single equilibrium droplet in a
two-dimensional binary fluid under shear flow. Despite the simplicity
of the model system it shows rich behavior, both droplet break-up and
droplet dissolution. We also suggest a new explanation of tip
streaming observed in our simulations.

Consider first an immiscible drop subjected to a shear flow. This
problem has been studied extensively since the original experiments by
Taylor \cite{taylor}. Experimental, theoretical and numerical results
are available in three dimensions \cite{rallison,stone,rallison2} and
theoretical \cite{richardson,buckmaster} and numerical \cite{ian1,ian2}
results in two dimensions. These approaches consider drops with a
singular interface and a conserved volume. The drops are deformed
by the shear flow while maintaining their volume. If the shear rate
exceeds a certain critical value, which depends on the volume of the
drop, the drop will break up. Conversely, for a given shear rate, there
exists a volume above which the drop is unstable. We shall denote this
volume $V_b$.

For a miscible binary mixture a similar break-up of droplets is observed
if the droplets are large. However there is now a second volume
scale $V_d$ which sets a {\em lower} limit to the drop size. $V_d$
corresponds to the minimum size of a nucleation seed. The
reason for the existence of $V_d$ lies in the free energy balance
between the favorable creation of separate phases in the
supersaturated mixture and the unfavorable creation of the interface
separating them. 
Note that, because when a shear is
applied a drop deforms and increases its surface length, $V_d$ will depend
on the shear rate. For shear rates with $V_b<V_d$ there are no stable
drops in the system.

To study these phenomena we use a model
developed by Orlandini {\it et.al.}\cite{binmix} for the isothermal
flow of a binary mixture. In this approach a free energy which
describes a binary fluid is chosen. The pressure tensor and chemical
potential calculated from this free energy are then included in a lattice
Boltzmann scheme for modeling fluid flow. The fluid obeys the
Navier-Stokes and convection-diffusion equations and comes to an
equilibrium corresponding to the minimum of the input free energy. The
coexistence curve is correct and interfaces are extended in space as
predicted by the Cahn-Hilliard theory. One advantage of this approach
is that the free energy and the chemical potential can easily be
calculated as local functions and compared to theoretical
predictions.  

In the next section of the paper we outline the thermodynamics of the
binary fluid and describe the extensions to the lattice Boltzmann
approach needed to treat shear flow. Results for the break-up of a
large droplet are presented in section three. In section four we obtain
an estimate for the volume $V_d$ below which small droplets dissolve
and discuss the effect of shear flow on the dissolution.  Section five
discusses tip streaming, the loss of material from the tips of the
deformed droplet, and summarizes the results of the paper.

\section{Method}
\noindent We simulate a binary fluid comprising two components A and B,
say. A--A and B--B interactions are zero but there is an A--B
repulsion $\lambda n_A n_B$ where $n_A$ and $n_B$ are the number
densities of A- and B-particles respectively. This system can be
described by the Landau free energy functional
\begin{equation}
\Psi = \int d{\bf r} (\psi(\varphi,n,T)
+ \frac{\kappa}{2} (\nabla \varphi)^2) \label{totf}
\end{equation}
where $T$ is the temperature, $n=n_A+n_B$, $\varphi=n_A-n_B$
and $\kappa$ is a measure for the interface free energy (surface
tension). The free energy density of the homogeneous system is
\begin{eqnarray}
\psi(\varphi,n,T)&=&\frac{\lambda n}{4}\left(1-\frac{\varphi^2}{n^2}\right) -Tn 
\nonumber\\
&&+\frac{T}{2}(n+\varphi) \ln\left(\frac{n+\varphi}{2}\right)
 +\frac{T}{2}(n-\varphi) \ln\left(\frac{n-\varphi}{2}\right). \label{bulkf}
\end{eqnarray}
The Navier-Stokes and convection-diffusion equations for the fluid are
simulated using a lattice Boltzmann approach described in detail in
\cite{binmix,lbref}. Here we restrict ourselves to a description of the
way in which shear flow was implemented.

We consider a
linear shear flow with velocity
\begin{equation}
\left( \begin{array}{c} u_x\\ u_y \end{array} \right) =
\left( \begin{array}{c} G y\\ 0 \end{array} \right) \label{ufield}
\end{equation}
where G is the shear rate. If the fluid is homogeneous it is expected
that eqn.(\ref{ufield}) describes the velocity field of the whole
fluid. Inserting a droplet will disturb the velocity field locally
but eqn.(\ref{ufield}) gives the far field solution.

To simulate the shear it is necessary to introduce boundary conditions
that force the flow. After each streaming step we replace the
collision step at the boundary by a step that defines the
variables of the lattice Boltzmann scheme to take values that correspond
to the required value of the velocity and densities.

Two different kinds of boundary conditions have been
implemented:
\begin{description}
\item{\it periodic:} The top and bottom edges of the lattice at $y=\pm
y_b$ are the boundaries and the velocity is constrained to be ${\bf
u}=(G y,0)$. The side boundaries have periodic
boundary conditions. This corresponds to a shear confined by two
moving walls acting on a periodic array of drops.
\item{\it forced:} All the edges of the lattice are forced to have
${\bf u}=(G y,0)$. This eliminates the effects of periodic images of
the drop. The local values of $n$ and $\varphi$ are replaced by their
mean value averaged over the boundary. These boundary conditions
generalize for more complicated forced flows, for example
hyperbolic shear flow.
\end{description}
For a homogeneous system both boundary conditions lead to the velocity
profile of eqn.(\ref{ufield}) to within machine accuracy. In the
presence of a drop both boundary conditions are expected to give the
same results for an infinite lattice. A comparison of the two
different boundary conditions therefore gives a measure of the effect
of the periodic images on the drop.

\begin{figure}
\begin{center}
\leavevmode
\epsfxsize=8cm \epsfbox{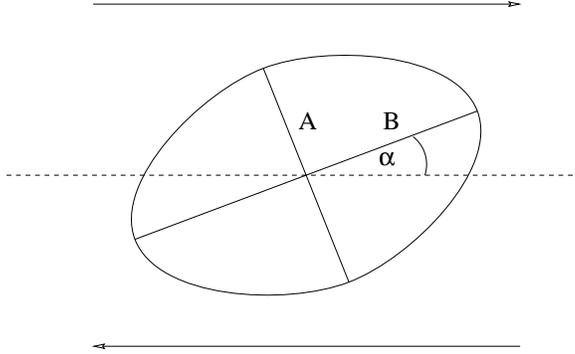}
\end{center}
\caption{Sketch of a deformed drop in a simple shear flow. For small
shear rates the drop has the form of an ellipse with axes 2A and
2B. The ellipse is inclined to the direction of the shear flow by a
shear strength dependent angle $\alpha$.}
\label{simpdrop}
\end{figure}

\begin{figure}
\begin{center}
\leavevmode
\epsfxsize=12cm \epsfbox{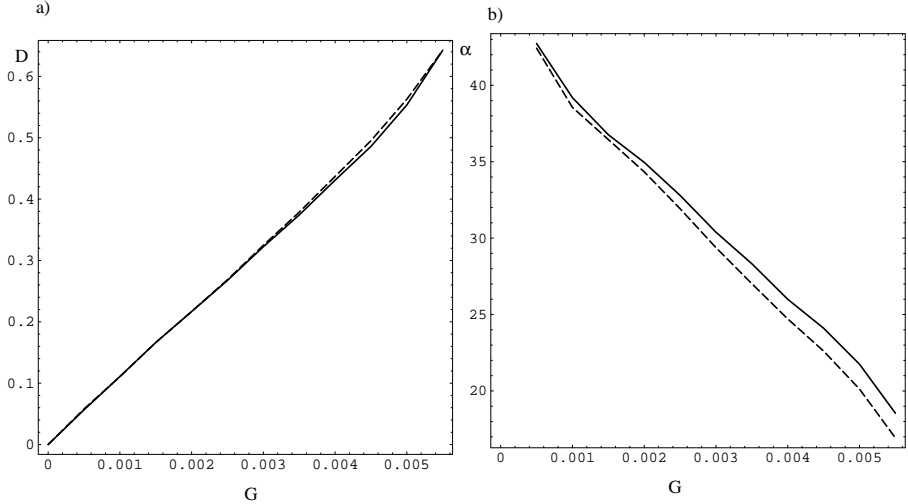}
\end{center}
\caption{a) Deformation of a drop $D$ and b) the tilting angle $\alpha$  
for periodic (---) and forced (- - -) boundary conditions
against the shear rate $G$. The undeformed drop has a radius of 8 lattice
spacings and the lattice size is 60x30.}
\label{comp}
\end{figure}

If a drop is placed in a shear flow it will be deformed by the forces
acting on it. The drop elongates and turns to lie at an angle
$\alpha$ to the flow until in the steady state the restoring force due
to the surface tension
balances the shear forces acting upon the drop. This situation is sketched
in figure \ref{simpdrop}. For small deformations the drop approximates
well to an ellipse and its deformation can be defined as 
\begin{equation}
D=\frac{A-B}{A+B}
\end{equation}
where $A$ and $B$ are the major and minor axes respectively.  For
drops of constant volume the deformation and inclination angle depend
only on a dimensionless quantity, the capillary number\cite{rallison}
\begin{equation}
\mbox{Ca}=\frac{\nu a G}{\sigma}
\end{equation}
where $\nu$ is the viscosity, $a$ the undeformed drop radius and
$\sigma $ the surface tension. Throughout this paper we take $\nu=1/6$
and $\sigma =0.046$ $(\kappa = 0.002)$. This corresponds to an
interface width $\approx 3$ lattice spacings.

\section{Break-up}
\noindent Typical results for large drops ($V\gg V_d$) are shown in figure
\ref{comp} where $D$ and $\alpha$ are plotted as function of the shear
rate. Results for forced and periodic boundary conditions are
compared. The results presented are for a lattice of size 60x30 with a
drop of initial radius 8. They were obtained by equilibrating the
fluid at each data point and then increasing the shear,
re-equilibrating to give the next point, and so on.

There is a linear dependence of the deformation on the shear rate for
small shear rates, followed by a more rapid deformation as the shear
increases, and finally break up. The different boundary conditions
lead only to small quantitative differences in the result.  The
results are in qualitative agreement with the results expected from
comparison with three-dimensional experiments.  For a drop of constant
volume and given ratio of the viscosity of the drop and the
surrounding fluid the deformation depends only on the capillary
number\cite{rallison}.  The drops studied here do not have a constant
volume and therefore deformation at break up should depend on the
parameters of the system but it is reasonable to expect only a weak
dependence. Indeed we find $D_b \approx 0.65$ for drops with
undeformed radii 8,13,20 in those cases where they break up.  

\begin{figure}
\begin{center}
\leavevmode
\epsfxsize=12.5cm \epsfbox{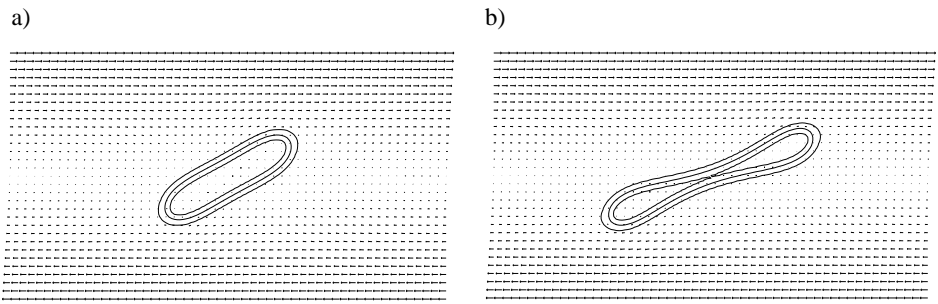}
\end{center}
\caption{ a) An equilibrium drop at $G=0.002246$, b) the same drop at
$G=0.002251$ in the first stage of breaking up. The drop in b) breaks
up into two drops of equal size at larger times. The arrows show the
velocity at every third lattice point in each direction.}
\label{dropbr}
\end{figure}

It is known\cite{stone} that the second
curvature is very important for the rupture of three-dimensional
drops. This mechanism does not exist in two-dimensional
systems. Therefore we felt it was important to check the existence of
the break up carefully, particularly as it is well known\cite{stone}
that a sudden change in shear strength can lead to rupture long
before the critical shear rate. We therefore performed careful
numerical simulations where we saved a stationary solution and then
increased the shear flow. If the drop ruptured instead of reaching a
stationary state we loaded the old configuration and increased the
shear rate by only half the previous amount. We iterated this step
until the increase in shear rate was smaller than a lower bound.  The
shear rate never grew larger than a previously rejected shear rate
showing that the break-up is not due to non-equilibrium
effects. We then decreased the shear rate to check that the system is 
in equilibrium at every point. No hysteresis effects were
observed showing that this is indeed the case.

It should be pointed out, however, that the rupture mechanism is very
different in two and three dimensions. Three dimensional drops break
up into two main drops and satellite drops\cite{stone}.  We observe
that two dimensional drops break up into two drops of equal
size as shown in figure \ref{dropbr}.

\begin{figure}
\begin{center}
\leavevmode
\epsfxsize=12.5cm \epsfbox{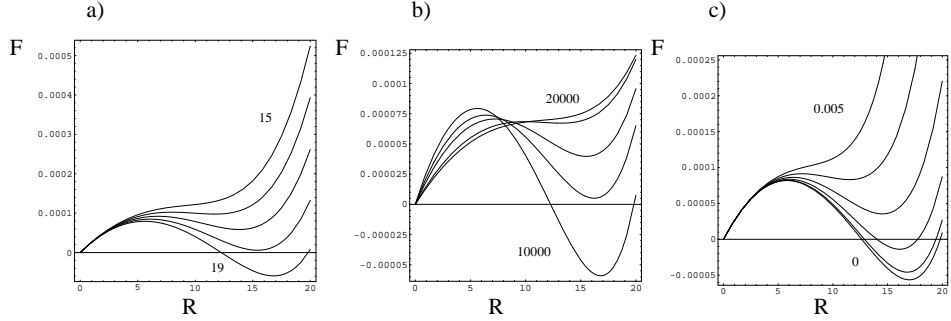}
\end{center}
\caption{The free energy plotted as function of the drop radius for a) a
lattice with 10000 points for concentrations that without surface effects
would correspond to drops of radii $R^0=$15,16,17,18 and 19; b) a
system with $R^0=19$ and lattice sizes 10000, 12000,
14000, 16000, 18000 and 20000; c) a system of size 10000 and
$R^0=19$ for shear rates $G=$0, 0.001, 0.002, 0.003, 0.004 and 0.005.}
\label{freegr}
\end{figure}

\section{Dissolution}
\noindent For given $n_A$, $n_B$, $T$, if a binary fluid lies in the
coexistence region it will separate into two phases with concentration
differences $\varphi_1$ and $\varphi_2$, say. Each phase occupies a
fraction of the total volume of the system which is determined by the
densities and temperature. For a finite system the interface between
the two phases results in a finite positive contribution to the free
energy and, if this is too large relative to the gain in free energy
due to the phase separation, the drop is unstable.

For the model simulated here it is possible to obtain an estimate of
the droplet volume below which dissolution will
occur. Ignoring the interface curvature the surface tension for an
interface orthogonal to the z-direction is\cite{widom}
\begin{equation}
\sigma = \kappa \int_{-\infty}^\infty \left(\frac{\partial
\varphi}{\partial z} \right)^2.
\end{equation}
Hence, using (\ref{totf}) the free energy of a drop of radius $R$
in a volume V is 
\begin{equation}
F=\pi R^2 \psi(\varphi_1,n,T) + (V-\pi R^2) 
\psi(\varphi_2,n,T)+ 2 \pi R \sigma.
\end{equation}
This should be minimized with respect to $R$, $\varphi_1$ and
$\varphi_2$ where one variable can be eliminated by the constraint 
\begin{equation}
\pi R^2 \varphi_1+ (V-\pi R^2) \varphi_2=\varphi. 
\end{equation}

In figure \ref{freegr}a the free energy, relative to that of a
homogeneous system, is plotted as a function of the drop radius for
different concentrations of A and B particles. The concentrations are
chosen such that without the surface effects drops of radii
15,16,17,18, and 19 would minimize the free energy on a lattice of
size 100x100. We will denote these radii by $R^0$.  Without the
surface free energy term there is only one minimum and this will
always be reached. With surface effects the homogeneous phase ($R=0$)
is always stable and a finite deviation from it is needed to reach the
global minimum. This corresponds to the metastability of some regions
of phase space where a finite nucleation barrier prevents immediate
phase separation.

The lowest free energy curve in figure \ref{freegr}a corresponds to
$R^0= 19$. The figure shows that the effect of the surface free energy
is to shift the minimum from 19 to about 16.7. For smaller $R^0$ the
minimum becomes a local minimum at $R^0\approx 18$. In systems with
fluctuations this will eventually lead to the drops
dissolving. However the lattice Boltzmann simulations reported here do
not include noise and a drop in a local minimum will be stable.  For
concentration ratios that lead to a graph that has no local minimum
($R^0\stackrel{>}{\sim} 15.5$) the drop will dissolve. However we observe
that the initial dynamics are very slow.

\begin{figure}
\begin{center}
\leavevmode
\epsfxsize=5cm \epsfbox{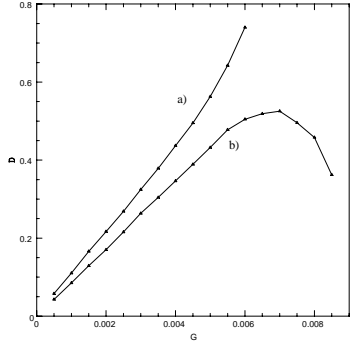}
\end{center}
\caption{Deformation as function of shear rate for a drop of radius 8
for systems of size a) 60x30, b) 70x40.  The drop in the smaller system
breaks up while the drop in the larger system dissolves. Because the
mass of the drop decreases strongly at high shear rates its
deformation decreases under the increase of shear before it completely
dissolves.}
\label{twodat}
\end{figure}

The radius below which drops dissolve depends not only on the size of
the drop but also on the total volume of the system. Results for
the concentration corresponding to $R^0=19$ for different system sizes
are shown in figure \ref{freegr}b. Droplets are less stable in a
larger system because more material from the drop is needed to change the
concentration outside the drop.

When shear is applied the droplets deform. They have a larger
interface so that the surface contributions are increased. For small
shear rates the dependence of the deformation on the shear rate is
approximately linear and we can use
\begin{equation}
D\approx 10 G R
\end{equation}
where $R$ is the radius of the undeformed drop
with the same volume. Approximating the shape of the deformed drop by
an ellipse the length of the interface is given by
\begin{equation}
L=\sqrt{\frac{1-D}{1+D}} E\left(-\frac{4D}{(1-D)^2}\right)
\end{equation}
where $E(k)$ denotes the complete elliptic integral of the second kind.
This result can be used to estimate the effect of shear flow on the free
energy. The results for $R^0=19$ for different shear rates are shown in
figure \ref{freegr}c. The minimum in the free energy vanishes
for high shear rates and no stable drops can exist. This corresponds
to the dissolution of a drop under shear. 

The effects predicted by the thermodynamic theory are observed in the
simulations.  In figure \ref{twodat} results for the simulation of a
drop of $R^0=8$ on lattices of size 60x30 and 70x20 are shown. For the
smaller lattice $V_b<V_d$ and the drop breaks up. For the larger
lattice the drop can exist to larger values of the shear, but loses
mass. Finally as the shear is increased further it dissolves. 

It is interesting to note that the theory predicts that every drop
will dissolve under shear in a sufficiently large system. That is
because the mass loss from the drop has to change the concentration
outside the drop in order to reach a new equilibrium. This can, however, be
very difficult to observe because there is a separation of time
scales. The time $t_D$ needed for diffusion to equilibrate the system
scales as $t_D\sim L^2$ where $L$ is the length of the system. This
result is altered under shear\cite{howard} and in the limit of
very large shear rates $t_D\sim L^\frac{3}{2}$. The time $t_F$
for the system to reach a steady flow and a new deformation of the
bubble scales as $t_F\sim L$. Therefore in large systems the time for the
deformation of the drop will be fast compared to the time for it to
dissolve.

\begin{figure}
\begin{center}
\leavevmode
\epsfxsize=12.5cm \epsfbox{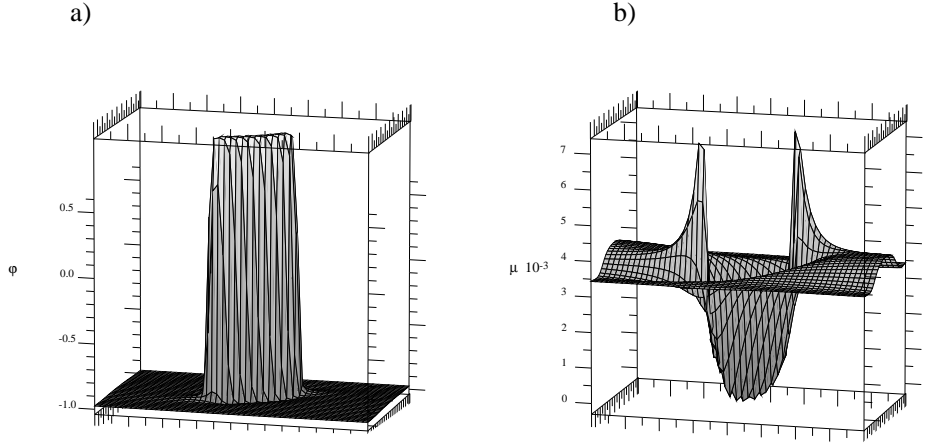}
\end{center}
\caption{a) The density difference $\varphi$ and b) the chemical
potential $\mu$ for a highly deformed drop.}
\label{tip}
\end{figure}

\newpage

\section{Discussion}
\noindent In our previous discussion we have assumed that
thermodynamic arguments for equilibrium can be carried over to the
stationary states of dynamic systems. This is, however, not
necessarily the case. We observe a deviation from this assumption in a
phenomenon which we will call ``tip-streaming''.  Mass is pulled from
the ends of the drop by the shear flow. The depletion of the
concentration in the drop leads to a reduction of the chemical
potential in the drop. The non-constant chemical potential leads to a
diffusion current acting in the direction of the chemical potential
gradient which returns material to the drop. Equilibrium is reached
when the diffusion into the drop balances the tip streaming. The
chemical potential in dynamic equilibrium is shown in
fig.\ref{tip}b. Note that this is a state of dynamic equilibrium
not thermodynamic equilibrium as $\mu$ is not constant.

For a small diffusion constant a large chemical potential difference
is needed for equilibrium and hence a large amount of mass is pulled
from the drop. This mechanism for the dissolving of a drop is distinct
from the free energy driven mechanisms explained earlier in the
paper. It depends on the diffusion constant which is a dynamical
quantity that does not enter the free energy. For infinite diffusion,
however, material is immediately returned to the drop and no
difference in the chemical potential is set up. A similar tip-streaming was
observed in experiments but was interpreted as a surfactant effect
\cite{stone}.\\

Halliday {\it et.al.} performed simulations on a droplet in a binary
fluid using a derivative of the the Gunstensen\cite{gunstensen}
algorithm. They obtain a similar break-up behavior but there are some
important differences in the results. In particular in the simulations
reported here all equilibrium drops are convex (see figure
\ref{dropbr}) whereas Halliday {\it et.al.}\cite{ian2} report stationary
drops that are constricted in the middle. They also observe smaller
inclination angles $\alpha$ at break up. 
The discrepancies warrant further investigation.
Another feature not reported by Halliday {\it et.al.}
is the dissolution of drops under shear. This may be due the fact that
they work in a region where the two components are more strongly
separated. Nevertheless small droplets should still dissolve. We
caution that the time-scale for dissolution can be much larger than that
for the deformation of the droplet to reach an equilibrium value.\\

Goldburg and Min\cite{goldburg} performed experiments on
nucleation in a binary mixture in the presence of shear. They observed
the vanishing of drops under the influence of shear as a sharp
transition. This transition may be interpreted as corresponding to the
point where $V_b=V_d$.\\

To conclude, in this paper we have shown that drops under increasing
shear flow can either break up or dissolve. We explained this behavior
with thermodynamic arguments. Tip-streaming was shown to lead to a
state of dynamic equilibrium for deformed droplets. 

\newpage
\noindent
{\bf Acknowledgments}\\
We thank E. Orlandini, C. Care, and I. Halliday for helpful
discussions. A.J.W. thanks J. Schlosser for providing visualization
software. J.M.Y. acknowledges support from the EPSRC, UK and NATO.

\nonumsection{References}

\end{document}